\documentclass[sigconf,9pt]{acmart}

\usepackage{todonotes}
\usepackage{listings}
\usepackage{xcolor}
\usepackage{hyperref}
\usepackage{caption}
\usepackage{graphicx}
\usepackage{array}

\AtBeginDocument{%
  \providecommand\BibTeX{{%
    \normalfont B\kern-0.5em{\scshape i\kern-0.25em b}\kern-0.8em\TeX}}}

\begin{document}

\title{WebAssembly Based Portable and Secure Sensor Interface for Internet of Things}

\author{Botong Ou}
\email{bou@purdue.edu}
\affiliation{%
\institution{Purdue University, West Lafayette}
}

\author{Baijian Yang}
\email{byang@purdue.edu}
\affiliation{%
\institution{Purdue University, West Lafayette}
}





\begin{abstract}

As the expansion of IoT connectivity continues to provide quality-of-life improvements around the world, they simultaneously introduce increasing privacy and security concerns. The lack of a clear definition in managing shared and protected access to IoT sensors offer channels by which devices can be compromised and sensitive data can be leaked. 
In recent years, WebAssembly has received considerable attention for its efficient application sandboxing suitable for embedded systems, making it a prime candidate for exploring a secure and portable sensor interface.

This paper introduces the first WebAssembly System Interface (WASI) extension offering a secure, portable, and low-footprint sandbox enabling multi-tenant access to sensor data across heterogeneous embedded devices.
The runtime extensions provide application memory isolation, ensure appropriate resource privileges by intercepting sensor access, and offer an MQTT-SN interface enabling in-network access control.
When targeting the WebAssembly byte-code with the associated runtime extensions implemented atop the Zephyr RTOS, our evaluation of sensor access indicates a latency overhead of 6\% with an additional memory footprint of 5\% when compared to native execution. As MQTT-SN requests are dominated by network delays, the WASI-SN implementation of MQTT-SN introduces less than 1\% additional latency with similar memory footprint.




\end{abstract}


\keywords{Embedded device, Device-to-Device Communication, Real-Time Systems, Security and Privacy, Webassembly}


\maketitle

\section{Introduction}

Internet of things (IoT) deployments have witnessed a rapid growth among domains ranging from home and industrial automation to agriculture and healthcare~\cite{rise_iot,predict_iot}.
Thousands of IoT devices (eg. sensors and actuators) are produced everyday~\cite{iot_trend1,iot_trend2}, and in their use they often need to engage in human activity or collect and report environmental data~\cite{iqbal2020wearable}. 
These functionalities necessitate networked deployments that enable interactive and collaborative access between users and IoT devices.

Ensuring adequate security across the resource-constrained devices that commonly comprise IoT systems is a challenging task, as
introducing security policies come at the cost of an increased  memory footprint, binary size, and runtime latency. As a result, most RTOS running on embedded devices are lightweight operating systems that discard certain security protections (e.g. Apparmor~\cite{gruenbacher2007apparmor} and Seccomp~\cite{corbet2009seccomp}), thereby leaving the devices vulnerable to attackers. This burden is passed onto the application developer, who must possess intimate embedded systems knowledge to ensure their application logic adheres to necessary security expectations~\cite{chen2018private}. Apriorit~\cite{Apriorit:2020} reports that most attacks targeting embedded systems can be categorized into software-based attacks~\cite{mullen2019assessment}, network-based attacks~\cite{stiawan2019investigating,andy2017attack} and side-channel attacks~\cite{gnad2019leaky,park2017using}. We focus on software and network-based attacks as they are considered to be both more common and easier to launch on embedded devices. 

To mitigate software-based attacks, we advocate for the adoption of the WebAssembly (Wasm) secure binary format~\cite{haas2017bringing}. As these attacks typically involve exploiting arbitrary memory access, an appropriate strategy to contain the exposed attack surface is to leverage software fault isolation (SFI). There have been some work to design a secure execution environment to contain applications' fault for IoT device. For exmaple, uSFI~\cite{aweke2018usfi} leverages rudimentary memory protection to isolate coupled application modules. S$\mu\mu$Vw~\cite{ammar2019s} introduced software based memory virtual mapping and use remote attestation to detect unexpected behavior. There is also a javascript based sandbox method~\cite{sahu2016securing} to secure IoT applications and prevent the faults from damaging the underlying systems. WebAssembly offers one such approach by establishing sandbox environments to isolate applications from the underlying embedded system. The WebAssembly runtime explicitly allocates memory to each independent module and intercepts accesses to ensure boundary and type checks. With the added benefits of the inherent portability of WebAssembly bytecode, embedded systems  developers are no longer required to be security experts. 

This work introduces WASI-SN, the first WebAssembly System Interface (WASI) extension offering Wasm modules secure access to on-board sensors. The runtime extensions expose a sensor capabilities API inspired by SmartThings~\cite{smartthings} to query, access, and configure attached sensors. Each request is managed by the shared runtime, enabling multi-tenant and safe access. Function-level signature verification is performed by the WebAssembly runtime to ensure that every application executing on the atomic sensor interface has the necessary privilege. Finally, WebAssembly alleviates limitations in portability across heterogeneous devices, as developers are no longer required to design platform-specific application implementations based on the associated deployment target. 




Network-based attacks commonly involve a lack of appropriate access control in the network stack; as such, we advocate for the upfront inclusion of access control in the application-layer network protocol. MQTT-SN is a natural choice for IoT networks spanning resource constrained embedded devices due to its (1) decentralized architecture, (2) reliance on UDP to minimize energy consumption, (3) support for many-to-many messaging, (4) decoupling of publishers and subscribers, and (5) minimal 2-byte header overhead. As such, our WebAssembly runtime implementation exposes an MQTT-SN API interface with the necessary access control primitives to guard against a large fraction of common network-based attacks. To achieve this, we propose a broadcast encryption scheme following the Wildcarded Identity-Based Encrytion (WKD-IBE) ~\cite{abdalla2011wildcarded} enabling data driven, end-to-end lightweight encryption while supporting flexible key delegation and revocation. Identity management is defined as a heirarchical structure to simplify reasoning through the key management process.


We offer an open-source implementation of WASI-SN with associated MQTT-SN extensions compatible with the Zephyr RTOS. A performance evaluation was performed on a testbed comprised of NRF52840-DK boards with Cortex-M4F@64 MHz MCU under the Armv7E-M architecture with IEEE 802.15.4 as the link and physical layer specification and an OpenThread thread network. The sensor access API introduced a 6\% overhead compared to native exectution; most of this latency was due to the fixed startup time required for the runtime itself to boot, upon which modules can be repeatedly executed. An additional memory footprint of 5\% was introduced during compilation. Leveraging the WASI-SN library within Wasm modules added a minimal memory footprint and and latency overhead of less than 1\% during runtime; this is due to request latency being primarily attributed to network delays.

Our contributions can be summarized as follows:
\begin{itemize}
    \item We offer a portable and generic sensor interface for commonly used embedded devices.
    \item We design a formal definition of sensor capabilities to constrain the sensor interface to necessary and valid accesses. 
    \item We advocate for WebAssembly as a new target for building embedded applications, and provide the first open-source implementation of a WebAssembly runtime with sensor interface extensions supporting Cortex-M processors. 
    \item We offer an MQTT-SN library built into the WebAssembly runtime that exposes the corresponding interface to WebAssembly modules.
    \item We introduce an access control scheme over MQTT-SN for different tenants with Wildcarded Identity-Based Encryption (WKD-IBE) and immediate key revocation.

\end{itemize}

\section{Background \& Motivation}











Webassembly (Wasm) is a new portable binary format designed to be available as a low-level compilation target. It is originally designed be replace Javascript in the browser in order to provide a lightweight sandbox environment to containerize code and prevent adversaries from harming the host, all while maintaining near-native performance. Given its use in browsers, Webassembly is designed to be compiled and instantiated quickly. Wasm enables the browser to utilize popular C/C++ libraries (e.g. libopus, libogg, libwebm) while maintaining good performance, standards compliance, and cross-browser usage~\cite{wasm_media}. 
More generally, Wasm offers portability and flexibility provided a compilation from a high level programming language(C/C++, Rust...) with an available runtime.

\textbf{Stack machine and static typing.} Wasm~\cite{haas2017bringing} is a well defined stacked virtual machine with static primitive types (i32, i63, f32 and f64) and pre-allocated linear memory, unlike the dynamic typing of Javascript. 
Consider the following Wasm code snippet exposing a function to multiply two values: 
\begin{center}
\begin{lstlisting}[caption={An example text format of Webassembly Module}]
(module
  (func $multiply (param $lhs i32)
  (param $rhs i32) (result i32)
    get_local $lhs
    get_local $rhs
    i32.mul)
  (export "multiply" (func $multiply))
)
\end{lstlisting}
\end{center}
All accesses are explicitly typed and only able to read from the sandboxed linear memory region. Operations are performed based on the most recent values pushed onto the stack. In the case of incorrect type or invalid memory access, the execution is halted to mitigate damage to the host environment.

\textbf{Control-flow integrity.} Wasm uses a function index space to enforce the control-flow integrity(CFI) of the program. Each Wasm program is composed of several structured Wasm modules, which serve as the unit of deployment, loading, and compilation. 
Each function call must specify a validated entry point to ensure a function can not point outside its table, and every call signatures is type checked. The runtime provides return address integrity by keeping a separate call stack which is not accessible by the Wasm module.

\textbf{Bounded memory access.} Each load and store operation is wrapped by a boundary check performed by the runtime. Before reading or writing to the pre-allocated linear memory, the runtime ensures the target address is within the boundary both before and after the operation. Combined with control-flow integrity, Wasm is thus immune to typical stack overflow and return-oriented programming attack(ROP) attacks, eliminating the need for common mitigation strategies including data execution prevention (DEP)~\cite{dessouky2018litehax} and stack smashing protection (SSP)~\cite{marco2019sspfa}.

\textbf{Minimal overhead compared to native.} The performance overhead of Wasm remains small when compared to native. For example, a "Hello World" program is 214 bytes in Wasm compared to 74 bytes in C. Although Wasm involves memory bounds checking, benchmark workload evaluation indicate that Wasm programs are typically no more than 1.5x the speed of native~\cite{Intel:2019}.

\textbf{Wasm outside the browser}
Although the primary use of Wasm has been to accompany Javascript within the browser to improve performance and security while maintaining portability, attention has recently shifted to its use outside the browser~\cite{wasm_outside1,wasm_outside2,hall2019execution}.
For example, Fastly~\cite{wasm_fastly} and Cloudflare~\cite{cloudflare} actively use Webassembly as the binary format to enable arbitrary programs hosted across Content-Delivery Networks. 
As of the writing of this paper, several runtimes have  been introduced supporting Wasm outside the browser, including Wasm3~\cite{wasm3} (C/C++, Rust, Golang) for devices including smartphones and microcontrollers, Wasmtime~\cite{wasmtime} (Rust and C++) for X86\_64 and AARCH64, and Wasm-micro-runtime (WAMR)~\cite{Intel:2019}, a low-footprint runtime specifically designed for embedded systems. WAMR supports various architecture including X86\_32/64, ARM/AARCH64, MIPS, and XTENSA, and is 100\% compliant to the W3C Wasm-MVP spec~\cite{w3c-spec}.
As WebAssembly offers a fast sandbox execution environment with the potential for SFI to containerize programs on embedded devices, WAMR offers a runtime that can be extended in support of sensor access and platform-agnostic access control.

\subsection{MQTT/MQTT-SN}

While the most practical wireless standard commonly used in ultra low energy IoT scenarios are radio technologies including Bluetooth Low Energy(BLE) and IEEE 802.15.4~\cite{brunner2020cross}, 
a proper application protocol can help maximize the quality of service provided by IoT networks. Examples include CoAP~\cite{bormann2012coap}, XMPP~\cite{seleznev2019industrial}, DDS~\cite{kang2012rdds}, MQTT~\cite{light2017mosquitto}, and MQTT-SN~\cite{stanford2013mqtt}. The Constrained Application Protocol(COAP) is designed as a resource-oriented interaction over UDP between users and end device resources. Extensible Messaging and Presence Protocol(XMPP) serves as message-oriented middleware based on XML language for exchanging structured but extensible data in real-time. Data Distribution Service(DDS) is a network middleware which implement publish and subscribe pattern to enable a machine-to-machine connection~\cite{garcia2017integration} between devices for distributed processing. Message Queuing Telemetry Transport(MQTT) is a TCP/IP based lightweight message protocol which supports pub/sub pattern for data exchange. Table \ref{table:protocols} depicts these protocols with respect to features including UDP transmission, decentralized architecture, decoupling sender/receiver, and multicast capabilities.

\begin{table}[t]
\caption{Comparison of Application Layer Protocols from several aspects}
\begin{tabular}{|c|c|c|c|c|}
\hline  
Protocols&UDP&Decentralized&Decouple&Communication\\
\hline  
COAP& $\surd$& $\times$&$\times$&1:1\\
\hline  
XMPP& $\times$& $\surd$&$\times$&M:N\\
\hline
DDS&$\surd$& $\surd$&$\times$&1:N\\
\hline
MQTT&$\times$& $\surd$&$\surd$&M:N\\
\hline
MQTT-SN&$\surd$& $\surd$&$\surd$&M:N\\
\hline
\end{tabular}
\label{table:protocols}
\end{table}

\begin{figure}[t] 
\centering 
\includegraphics[width=\linewidth]{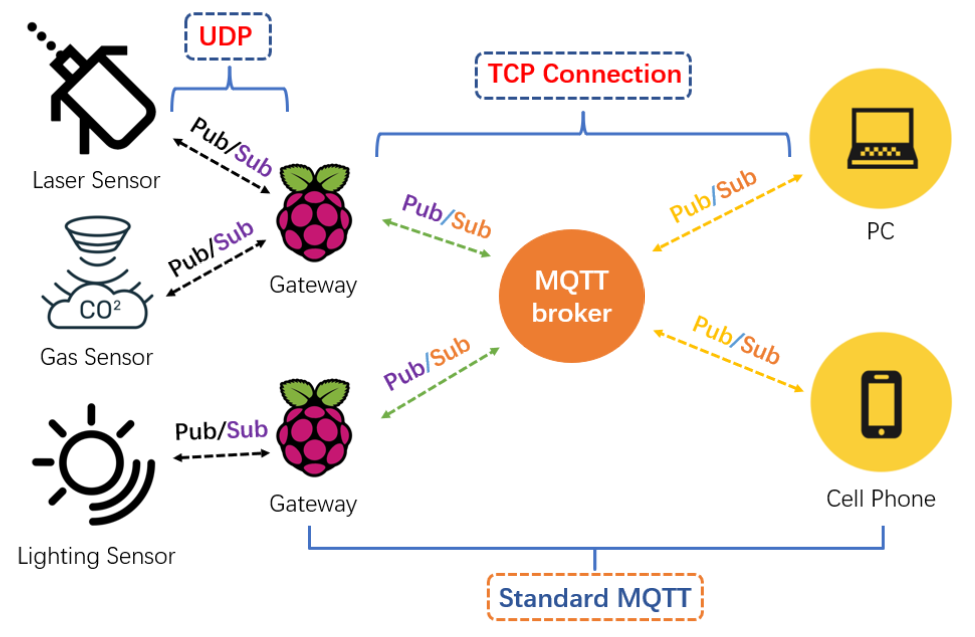}
\caption{Architecture of MQTT-SN (Hybrid Layout).} 
\label{MQTT-SN}
\end{figure}

MQTT-SN is a lightweight application level protocol based on the traditional MQTT protocol and is designed for wireless sensor networks (WSN)~\cite{park2017wireless}. 
The main entities forming an MQTT network are the \textbf{brokers} and \textbf{clients}. The broker is responsible for maintaining the message queue containing data received from various client publishers and distributes them to the correct client subscribers. Clients serve as the end user that either publish data to a broker or subscribe to receive published messages.
Connections are established by \textbf{topic}, a unique identifier for a series of messages. 
A key benefit of the pub-sub scheme is the decoupling of clients while maintaining service assurance. 
Then the broker collects messages and distributes them to all subscribed clients. The standard MQTT architecture~\cite{steger2017cesar} is shown in the right part of the Figure \ref{MQTT-SN}.

\begin{figure}[ht] 
\centering 
\includegraphics[width=\linewidth]{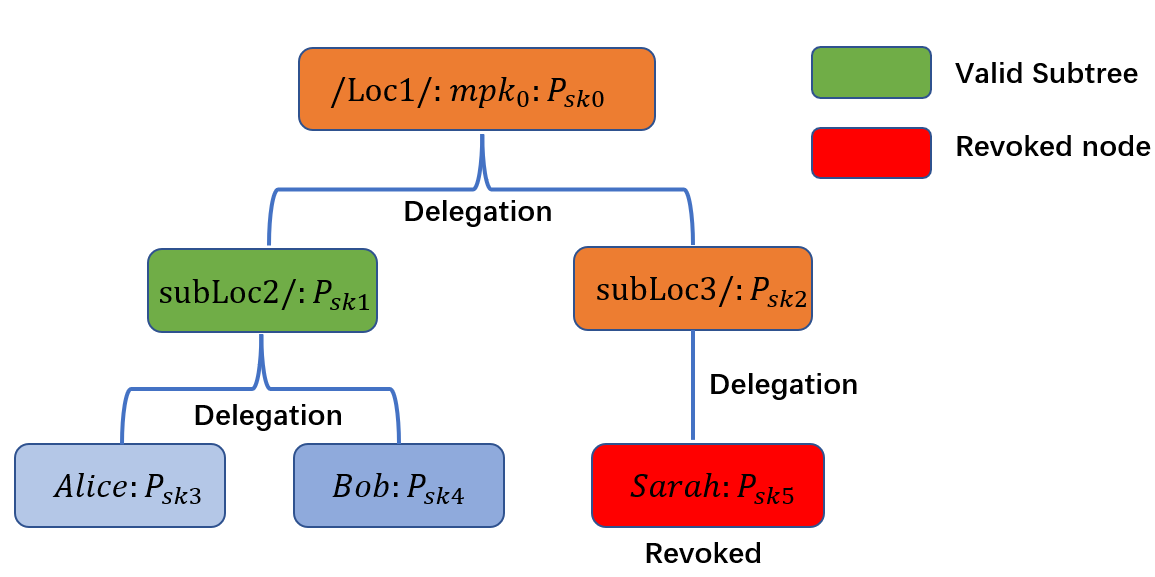}
\caption{Key management of Complete Tree. When Sarah has her key revoked, subLoc2/ is the remaining subtree root covering all the remaining valid nodes} 
\label{Fig.main2}
\end{figure}

MQTT poses certain limitations that prevent its deployment into embedded systems, thus motivating the design of MQTT-SN. 
The key inclusion in the MQTT-SN architecture is the \textbf{gateway} component to interface resource-consrtained clients with the MQTT broker.
First, the TCP/IP network stack is well known to not be an energy-efficient means to maintain long-term sessions on resource-contrained devices. As such, devices in MQTT-SN communication with a gateway through a UDP connection that reduces the overall energy consumption; the gateway then interfaces with the broker through standard TCP. 
Second, the MQTT topic definition uses a long string format that makes it impractical over IEEE 802.15.4 radios. 
MQTT-SN introduces a 16-bit length integer to replace the arbitrary string topic name. The end device then uses a new \textit{register} command to allow the gateway to map the 16-bit integer to a valid MQTT topic. 
Finally, MQTT-SN allows clients to enter a sleep mode while preserving active connections. The gateway buffers downlink messages while clients sleep, and are provided the messages upon reawakening.
We selected the hybrid architecture, as depicted in Figure \ref{Fig.main2}, in our implementation to eliminate single points of failure.

\section{Design Challenges \& Goals}
There are several challenges of directly extending WebAssembly to access sensors on  embedded systems with MQTT-SN for application-level networking.

\textbf{Interactive embedded systems.} Although embedded systems can typically run a given tasks for extended periods, updating an application without physically reflashing the device is a nontrivial challenge. The inclusion of a dynamic runtime with networked control can allow developers to deploy applications remotely and without physically accessing the device.

\textbf{Define a clear standard.} There is no obvious standard for how to interface sensor access to their pertinent services. Certain RTOS (e.g. Zephyr\cite{mideus2020risc}, VxWorks\cite{casini2016support}) follow the libc interface design to wrap low-level functions provided by platform-specific SDKs. Unfortunately, the libc standard does not cover interfacing sensor access, yielding platform specific implementations that vary with the RTOS. This restricts portability of application binaries across devices.

\textbf{Minimize overhead.} The tradeoff of additional latency and memory footprint commonly limits the inclusion of security mechanisms on RTOS. Wasm can help fill the gap left by DEP and SEP by providing a self-contained environment.

\textbf{Scalable fine-grained access control.} Wasm modules running on an embedded device must have access to the underlying service and APIs without coarse provisioning that assigns more than needed. This necessitates function-level fine-grained access control, both to device resources and MQTT-SN topics. Furthermore, a signature representing the identity Wasm module must be verifiable by the system. Finally, given the potential for large deployments mandates a scalable access control scheme. 

\textbf{Secure and encrypted session channel.} MQTT-SN only provides a TLS/DTLS secure session channel between clients and the broker. There is a need for protection mechanisms to ensure only the appropriate clients receive and can decode the published data.

\textbf{Untrusted MQTT broker.} The traditional MQTT-SN design assumes a trusted broker. If compromised, any TLS between client and broker is no longer safe, and the published messages may be leaked. The provided access control scheme should be designed without involving the broker as part of the trust anchor.

\section{WASI Extended Interface}


\begin{figure}[t] 
\centering 
\includegraphics[width=\linewidth]{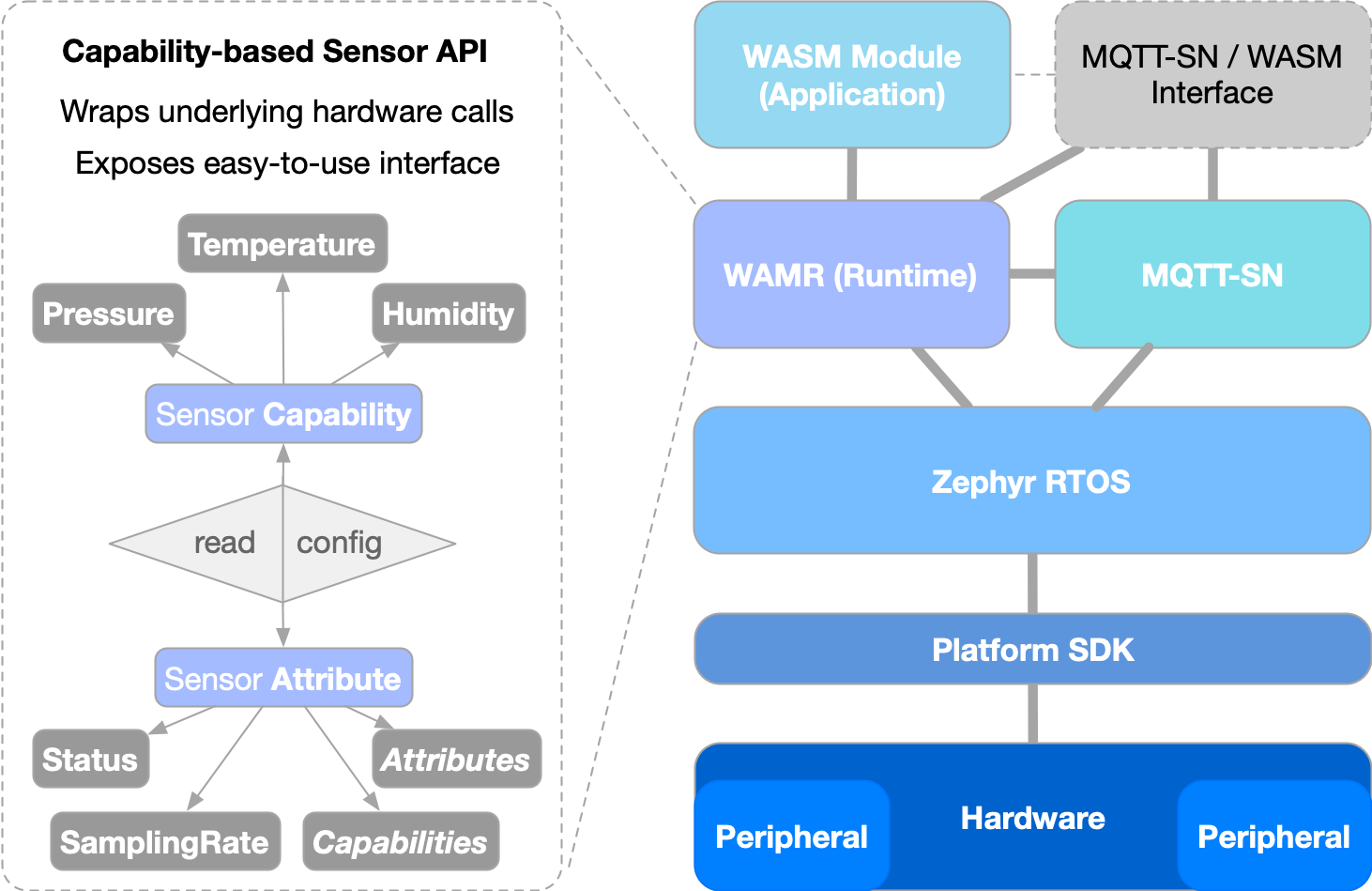}
\caption{Sensor interfaces illustration (left) to configure and access device sensors with an application stack diagram (right).} 
\label{Fig.sensor-interface}
\end{figure}

The proposed sensor interface design follows the official WebAssembly System Interface (WASI) offered by the W3C group. WASI is an API standard following the POSIX specification designed to provide Wasm module with the access to operating-system-like features such as filesystems, clocks, and entropy generators. In short, WASI provides a generic wrapper over the native system interfaces to wrap them into the interfaces for the WASM module with a capability check. The design pattern is reminiscent of Samsung's SmartThings, which define sensors based on their available capabilities (e.g. an audio controller supports setVolume, turnUp and turnDown). Unlike SmartThings, which focuses relies on proxy representations of devices in the cloud, our generic capability based sensor interface exposes device peripherals directly to local Wasm applications.
Accessing sensors thus becomes similar to accessing files in an operating system. A capability exposed by the device is mapped with a corresponding Wasm interface. 

When a wasm module executes a function call, the runtime intercepts the call and redirects it to the appropriate callee to begin execution. If a module calls functions that are not implemented by the runtime itself, but instead exposed by the operating system or other libraries, the control flow is validated and directed to the appropriate source. The entire execution procedure occurs within the linear memory pre-allocated to the Wasm module. For example, to access a sensor, the Wasm runtime must wrap the underlying implementation of either the RTOS or platform's SDK into the interface that are exposed to the Wasm modules. The system stack and control flow of the program is depicted in Figure \ref{Fig.sensor-interface}.

\subsection{Sensor Interface Primitives}

Table \ref{table:sensorinterface} outlines the sensor interface primitives. A sensor is defined by a set of \textbf{capabilities} and \textbf{attributes}. A capability represents underlying sensor functionality. For example, a BME280 may provide temperature, humidity, and pressure capabilities, while a CCS811 sensor may offer $CO_2$ and Volatile Organic Compounds(VOC) capabilities. Attributes represent sensor state and configuration, such as the active sampling rate or the current value of an LCD display. To query for these values, every sensor has two universal attributes, the \textbf{capabilities} attribute dictating a sensor's list of capabilities, and an \textbf{attributes} attribute dictates the sensor's list of attributes. Each capability and attribute is explicitly granted read and/or write access to a module via the in-network access control described in Section 5. As this interface is meant to be used in Wasm (i.e. compiled from a higher-level language), the interface primitives are low-level and generic, thus allowing ease in translation from a higher-level language.

\textbf{getSensors}: We first provide an interface to get the list of available sensors on the embedded device. This interface will read the global configuration of the device to see which sensors are attached. Invoking \textbf{getSensors} returns the list of sensors IDs associated with attached sensors, such as BME280, CCS811 or an LCD.

\begin{table*}[t!]
\caption{Summary of sensor interface primitives}
\begin{tabular}{|c|c|c|c|}
\hline  
Interface&Parameters&Return Value&Functionality\\
\hline  
\textbf{getSensors}&\textit{None}&String array&Fetch list of available sensor IDs\\
\hline  
\textbf{turnOn}&Sensor ID&Boolean&Initialize the sensor if needed\\
\hline  
\textbf{turnOff}&Sensor ID&Boolean&Deinitialize the sensor\\
\hline
\textbf{config}&Sensor ID, attribute, value&Boolean&Update a sensor's attribute (e.g. sampling rate)\\
\hline
\textbf{read}&Sensor ID, (capability | attribute), buffer, length&Boolean&Update buffer with current sensor or attribute value\\
\hline
\end{tabular}
\label{table:sensorinterface}
\end{table*}

\textbf{turnOn/Off}: This interface is used to initiate and deinitialize a sensor for subsequent access from a module. The runtime intercepts these accesses to ensure appropriate usage and execute the underlying SDK operations. For example, \textit{turnOn("BME280")} may initialize the sensor (if not already initialized by the runtime) and, if needed, set the sampling rate to 1 second. To avoid race conditions of multiple modules attempting to modify the same device, \textit{turnOff} can be optionally reserved for the runtime, or may be stubbed out with a runtime check that first ensures that no modules are actively registered to the specified sensor, such as in our implementation.

\textbf{config}: Updating attributes of a sensor are performed with the \textit{config} interface. 
This interface can be used to configure sampling rates and other relevant state variables. 
Configurations are provided as key-value pairs to indicate the specified attribute values. For example, config("BME280", "SamplingRate", 1000000000) indicates that the BME280 sensor should report the humidity value every second, thus overwriting the previously specified (e.g. default) sampling interval. The runtime performs validation of module-specified configuration values to ensure they do not violate the device state. By first exposing a sensor's attribute and then providing a module access to these values, a developer is offered the ability to configure sensor state. Each attribute read/write permission is explicitly granted to each module. This interface is intentionally generic, thus allowing a developer flexible access into controlling sensor state.

\textbf{read}: The read primitive retrieves the active capability or attribute values based on the specified parameter (i.e. sensor or configuration requested). For example, reading from the BME280 sensor with humidity, temperature, and pressure fetches the current sensor value. Similarly, sampling rates and other state configurations can also be obtained. The list of available capabilities and attributes can be obtained by performing a read on the two universal attributes, e.g. \textit{read("BME280","capabilities")} and \textit{read("BME280","attributes")}. 

\textbf{state}: This state attribute is a unique and universal attribute indicating the operating state of a sensor. 
The current valid state enum values includes \textit{Alive}, \textit{Stopped}, \textit{Working} and \textit{Sleeping}. 
State updates are reserved for the runtime, which maintains the active state to be read by running modules. When a sensor is turned on, the runtime initializes the sensor and updates the state to \textit{Alive}. Similarly when a read or config update is performed, the sensor state is briefly updated to the \textit{Working} state to indicate that the sensor is actively being used. All accesses by modules are buffered by the runtime and subsequently performed.


\subsubsection{Sensor Interface Example}
The following example indicates a C program that connects to a BME280 sensor and accesses the current humidity, which is compiled to a similar Wasm program:

\begin{lstlisting}[caption={An example C program that compiles to Wasm using the sensor interface}]
int main(){
  turnOn("BME280");
  int rate = 1000000000; //ns
  config("BME280", "SampRate", rate);
  char buf[] = {0,0,0,0};
  read("BME280", "humidity", buf, 4);
} 
\end{lstlisting}

%
\subsection{Network Interface Primitives}
Firstly, we show the network interface in the following table as a summart:
\begin{table*}[t!]
\caption{Summary of network interface primitives}
\begin{tabular}{|c|c|c|c|}
\hline  
Interfaces&Parameters&Return Value&Functionality\\
\hline  
\textbf{start}&Port&Bonded port&Start MQTT-SN client\\
\hline  
\textbf{stop}&\textit{None}&Boolean&Stop MQTT-SN client\\
\hline
\textbf{searchGW}&Broadcast Address, Port, MaxHops&Gateway address&Identify gateway address via broadcast\\
\hline
\textbf{connect}&Client ID, KeepAlive, Gateway, Port&Boolean& Connect to gateway with session duration\\
\hline
\textbf{register}&topicName&TopicID&Register topics in broker and get mapped ID\\
\hline
\textbf{publish}&topicID, QoS, Payload&Boolean&Publish data to a topic at a QoS level\\
\hline
\textbf{subscribe}&topicID, QoS&Boolean&Subscribe a module to a given topic\\
\hline
\textbf{hasMessage}&topicID&Boolean&Indicates if runtime has available message\\
\hline
\textbf{getMessage}&topicID&Byte array&Returns next published message\\
\hline
\textbf{sleep}&length (optional)&Boolean&Temporarily buffer messages at the gateway\\
\hline
\textbf{awake}&\textit{None}&Boolean&Resume receiving messages\\
\hline
\end{tabular}
\label{table:mqttsninterface}
\end{table*}
The MQTT-SN network interface exposed in the Wasm runtime follow the same specification as MQTT-SN itself, thus easing the burden of translating higher-level languages to the compiled Wasm target. Interface requests are translated into the equivalent request and implemented by the runtime. Table \ref{table:mqttsninterface} summarizes this interface, which includes the following:

\textbf{searchGW:} This API is used to find out the address of MQTT-SN gateway by multicast a message to a multicast to a IPv6 address which is listening by the gateway at a specific port. Also it will take hop limits as the parameter to indicate the maximum number of hops allowed for a given message.

\textbf{start:} This API simply starts an MQTT-SN client on a specific UDP port of the end device. The future network messages will be sent on this port and this should be the first interface called before any network operation.

\textbf{connect:} This allows a client to connect to the gateway with a given address (either pre-known or determined by the \textit{searchGW} interface). By default, the connection will have a 30 second keep-alive time. Each connect message waits at most 10 seconds before receiving the ACK from the gateway. Given the UDP nature of MQTT-SN, the maximum number of re-transmissions is three, after which a client will stop sending additional connect messages.

\textbf{register:} This interface is used to register a topic to the broker. Before clients publish their data to the broker, they must register the topic name for the broker to establish the mapping between the topic name and the 16bits MQTT-SN topic ID. The broker first sends an acknowledgement indicating a healthy connection, and then sends a message containing the converted topic ID for the client and gateway to use.

\textbf{publish/subscribe:} These two interfaces are used to publish and subscribe to a specific topic. When using the publish interface to publish data, the data must be converted to a byte array to be sent to the topic ID mapping, which is translated to the MQTT topic name. Similar to the MQTT-SN interface, the user can specify the desired quality-of-service from ${0,1,2}$. A value of 0 indicates that the client only publishes data once, without checking for a successfuly delivery. A value of 1 indicates that the publisher will continuously publish until it receives a PUBACK. Finally, a value of 2 indicates a 2-step handshake between the broker and client to ensure idempotent publishing; each message is guaranteed to be published only once, assuming a valid connection is available. When using Subscribe API to subscribe to a topic, it can use either the topic name or topic ID to subscribe to the specified topic. It also supports quality-of-service configurations akin to the publish interface.

\textbf{hasMessage/getMessage:} Due to the structured control flow of Wasm, the runtime validates and registers its own callback for published MQTT messages. The \textit{hasMessage} interface allows a module to query the runtime as to whether a message is available, upon which it can access the message via the \textit{getMessage} API, given the module request is valid and registered with appropriate access control.

\textbf{sleep/awake:} The sleep command pauses the MQTT-SN client on the device and saves the connection state within the runtime. The awake command will resume the MQTT-SN connections based on the saved state, allowing a device to sleep for extended periods of time while maintaining active MQTT-SN connections. The gateway will buffer messages that are published during sleep, and forward them to the client upon waking.

\subsubsection{Network Interface Example}

The following example illustrates a C program (which compiles to an equivalent Wasm module) that starts an MQTT-SN client, connects to the gateway, registers a topic, publishes a value, and disconnects. 

\begin{lstlisting}[caption={An example C program that compiles to Wasm using the MQTT-SN network interface}]
int main(){
  int KeepAlive = 30; //sec
  string ClientID="WASM"
  string IP="fdde:ad00..."
  int Port = 47193;
  start(1000);
  connect(ClientID,KeepAlive,IP,Port);
  int topicId = register("humidity");
  publish(topicId, 2, data);
  disconnect();
} 
\end{lstlisting}

%

\section{Access Control}
Providing both the sensor interface and MQTT-SN interface allows Wasm applications to access the on-board sensors and connect with remote devices. However, as most microcontroller and RTOS do not provide a substantial access control scheme to restrict the application behavior, we introduce a WKD-IBE based encryption scheme and an access control over MQTT-SN. The benefits of wildcard-based encryption are fairly straightforward: pattern delegations can be chained and grouped, thus naturally lending itself to the many-to-many messaging model of MQTT-SN. 
\subsection{WKD-IBE}
WKD-IBE stands for wildcard identity based encryption. It's an encryption scheme which uses the \textit{patterns} to encrypt messages. Instead of using public/private key pairs related to different entity, the key used in WKD-IBE is bonded to the patterns. A pattern is defined as the following:$P=(P_1,...,P_l)\in (\{0,1\}^*\bigcup \{*\})^l$ where $l\leq L$. The $*$ is a special wildcard symbol and $L$ is the maximal depth of WKD-IBE scheme. $P$ represents a pattern vector comprised of $l$ components $P_i, 1\leq i\leq l$ 
A pattern $P^*$ and $P'$ matches with each other denoted as $P^* \in_* P'$ only when for every $i\in \{1,...,l\}, P^*(i)=\{*\}$ or $P^*(i)=P'(i)$. To be specific, only when every element in patter $P^*$ equals to either special symbol $*$ indicating any or the element in $P'$ at the same location, does it mean $P^*$ matches with $P'$. The reverse is not necessarily true, i.e. $P'$ matches with $P^*$.

WKD-IBE specifies key for each pattern. A key for pattern $P^*$ can decrypt the message which is encrypted by the pattern $P^*$. The same key can also decrypt all messages encrypted by the pattern $P'$ if $P^*$ matches $P'$. If there is a need to generate a new key for pattern $P'$, the key for $P^*$ can also be used to the key for $P'$ when it matches with $P'$. In general, the procedures to use WKD-IBE scheme is the following:

\textbf{Setup}: During the setup, the root identity randomly choose $params=g, h_0,...h_L\in \{0,1\}^*$ to be the parameters for deriving the master public key and private key. The master public key will be $mpk\leftarrow (g^\alpha,params)$ where $\alpha \in \mathbb{Z_p}$ is a prime number. As for the master secret key, it will be derived by $msk\leftarrow {h_0^\alpha}$

\textbf{Key Derivation}: Next, keys for specific patterns are derived. The key derivation function takes $params$,$secret$ $key$ and $P=(P_1,P_2,...,P_l)$ which is the target pattern as the input to derive the secret key $P_{sk}$ for the pattern $P$. For the $secret$ $key$, it can be either the $msk$ derived at the \textbf{Setup} stage or the derived $P_{sk}$ to derive a new key $P'_{sk}$ for a matched new pattern $P'$. For more details on key derivation, please refer to ~\cite{abdalla2011wildcarded}.

\textbf{Encryption}: To encrypt a message $m$, the encryption method requires the $params$ and pattern $P$ as input. For example, if a message $m$ is encrypted for the pattern $P'=(P'_1,...,P'_l)\in_* P$, the sender (who already has the $P_{sk}$) uses $mpk=(params,g^\alpha)$ to compute several ciphertext blocks to assemble into the final ciphertext. The several blocks are computed as:
\begin{enumerate}
    \item $C_0=g^t$ where $t$ is a random chosen prime number
    \item $C_i=H_i(P'_i)^t$ where $i\in \{1,...,l\}$. $H_i$ is the $L$ independent random oracles mapping $\{0,1\}^*\rightarrow \mathbb{G}$ where $\mathbb{G}$ is group of prime order $p$
    \item $C_i=h^t_i$ for every $i\in \{l+1,...,L\}$
    \item $C_{L+1}=m\times \hat{e}(g_1,h_0)^t$
\end{enumerate}
The final ciphertext is assembled as $C=(C_0,...,C_{L+1})$.

\textbf{Decryption}: When decrypting a message encrypted by the pattern $P$, the receiver should use either the private key $P_{sk}$ or the derived key $P'_{sk}$.

The basic encryption scheme as described is used to ensure that all messages are end-to-end encrypted, and only clients granted access are able to decrypt published messages. However, WKD-IBE does not address the problem of key delegation; as such, a hierarchical key structure is needed to delegate and revoke keys.

\begin{table*}[ht]
 \newcolumntype{C}{>{\raggedright\let\newline\\\arraybackslash\hspace{0pt}}m{0.3\linewidth} }
 \newcolumntype{D}{>{\raggedright\arraybackslash} m{0.7\linewidth} }
 \resizebox{\linewidth}{!}{
  \begin{tabular}{CD}
   \toprule
   \textbf{Resource} & \textbf{Hierarchy Structure} \\ 
   \midrule
   \midrule
   Capabilities of BME280
   & /resources/<device ID>/BME280/<Capabilities> \\
   \midrule
   Attributes of BME280
   & /resources/<device ID>/BME280/<Attributes> \\
   \midrule

   Requester UII 
   & /entity/<location>*/<requester ID>/<delegated User>\\
   Request Payload
   & <Interfaces|params>;<Requester>;<Requester's Public Key> \\
   \bottomrule
  \end{tabular}
 }
 \caption{Example access control resource and requester hierarchy}
 \label{tab:hierarchystruct}
\end{table*}

\subsection{Hierarchy of Resources and Requesters}
WKD-IBE is designed to serve a single hierarchy for key delegation. The intended use case is to delegate wildcard keys such as in email addresses catchalls. When applying WKD-IBE to embedded systems, the structure of embedded resources can represent a similar structure to email addresses. For a remote requester, resources can be represented in a hierarchy composed of several domains. For example, a temperature sensor located in a factory could be represented by $A$ in Table \ref{tab:hierarchystruct}, which jointly serves as the URI to describe the remote resources.

\begin{table*}[t]
 \newcolumntype{C}{>{\raggedright\let\newline\\\arraybackslash\hspace{0pt}}m{0.3\linewidth} }
 \newcolumntype{D}{>{\raggedright\arraybackslash} m{0.7\linewidth} }
 \resizebox{\linewidth}{!}{
  \begin{tabular}{CD}
   \toprule
   \textbf{Topics} & \textbf{Topic Format} \\ 
   \midrule
   \midrule
   Request capability of BME280
   & /resources/<device ID>/BME280/<Capabilities>/request \\
   \midrule
   Request attribute of BME280
   & /resources/<device ID>/BME280/<Attributes>/request \\
   \midrule
   Topics for issuing key from BME280:
   &/resources/<device ID>/BME280/<Capabilities>|<Attributes>/issue\\
   \midrule
   Topics for receiving return value of interfaces from BME280:
   &/resources/<device ID>/BME280/<Capabilities>|<Attributes>/ret\\
   \bottomrule
  \end{tabular}
 }
 \caption{MQTT-SN topics used in the example described in Table \ref{tab:hierarchystruct}}
 \label{tab:mqttsntopic}
\end{table*}

To better manage different keys for different requesters and enable revoking keys for a subgroup when necessary, a hierarchy to represent the uniform identity identifier(UII) is also needed. A structured user identity is also introduced as the following. For example, a user named Alice needs the temperature value from the BME280 sensor at a local factory. The structured identity of Alice then becomes $/entity/location/Alice/NULL$. The domain between the $entity$ and the name of the user is the geometry location of the requester(The name of the user is controlled by the requester). The \textit{delegated User} domain represents the target entity granted permission by $Alice$ (NULL indicates no delegation).

After creating hierarchies of both the resources and the requesters, a network level key exchange scheme is required to let the remote embedded device delegate keys to authorized groups.

\subsection{Key Establishment}
In most cases it is advantageous to not require trust in the broker; ideally, only authorized end-users should have access to messages. As such, the decryption key of the WKD-IBE scheme should be only accessible to the authorized group. In this case, we build a key exchange scheme similar to what TLS does in the transport layer over the MQTT-SN application layer protocol. We first define the key establishment for different capabilities and attributes that a remote user requests.

Let us consider the BME280 sensor as an exmaple. BME280 is able to sense the temperature, humidity and pressure of the environment. Therefore, the remote resource URI of BME280 will be \\$/resources/device01/BME280/temperature$ for temperature capability. When the end device is capturing the temperature value and publishing to the broker, the value content should be encrypted to be only accessible by the remote user. The key used to encrypt the value is the master public key $mpk$ derived by WKD-IBE. However, for the pattern used in encryption and decryption of WKD-IBE, since it should be only accessible for authorized group, the key derivation should also include both the URI and the structured identity as the pattern to be the input parameter.

We assume there are three identities which are $../Loc1/Alice$, $../Loc1/Bob$ and $../Loc2/Sarah$ want to access the temperate attribute of the sensor BME280. And we also assume people in $Loc1$ are authorized to access the data. In this case, the pattern used to encrypt the sensor data is the following: $P=(P_1,P_2)$ where $P_1=/entity/Loc1/*$ and $P_2$=\textit{/resources/device01/BME280/temperature}. The together the pattern $P$, the master public key $mpk$ owned by end device, the data $m$ of the sensor value is able to be encrypted. By far, the encryption stage has finished.

When generating the decryption key corresponding to the pattern $P$, the key derivation is different between $Alice$ and $Bob$. To derive the decryption key for $Alice$, the device use master secret key $msk$ and pattern $P'=(P'_1,P'_2)$ where $P'_1=/entity/Loc1/Alice$ and $P'_2=P_2$ to generate the decryption key $P'_{sk}$. When deriving the key for $Bob$, the pattern changes to $P''=(P''_1,P''_2)$ where $P''_1=/entity/Loc1/Bob$ and $P''_2=P_2$ to derive the key $P''_{sk}$. $P'_{sk}$ and $P''_{sk}$ are both able to decrypt the message encrypted by the pattern $P$ because $P'_1\in_* P_1$ and $P''_1\in_* P_1$, indicating that both patterns match the encryption pattern.

\subsection{Key Exchange}

The decryption key establishment occurs when the end device receives the request from remote user asking for permission to access one of the sensor values. Here we introduce TLS-like methods over MQTT-SN to establish the key exchange. The user needs to publish a request asking for authorization with the data decryption key of a specific URI. We define the topic that the user should publish the request to as the \textit{Request Payload} shown in table \ref{tab:hierarchystruct}. All of the domain before the $key$ represent the URI of the attribute. The requester inserts the requested capability into the message payload, which is composed of several fields: interfaces, capabilities, and attributes which the requester wants to execute on the device, the user's identity $I_U$ following the previously described definition syntax, and the public key of the requester $U_{pk}$. The full payload is structured similar to $read;/entity/Loc1/Alice;Alice_{pk}$ and published onto the topic $/resources/<device ID>/BME280/temperature/request$ if a user $Alice$ wants to request the temperature value of the $BME280$ sensor. If the user wants to have $config$ access, the only difference in the payload is to replace $read$ with $config|sensorID,attribute$. 

After the requester has generated the payload, and in order to not involve the broker as the trust anchor in the framework, the payload content is encrypted by the device's public key $D_{pk}$, which is only accessible by the device itself. Here we assume both the public keys of the requester and the device are generated by a trusted third party like the PKG center (the keys are only used to ask and grant permissions). Once the data is encrypted, it's published to the broker and received by the device.  On the device side, to receive the permission request, it should be subscribed to the same topic. To enable the availability of the framework, the device should always subscribe to the main topic \textit{/resources/device01/*}. Whenever a permission request is published by the user to any sensor attributes under the $device01$, the end device can always receive the request. Upon receiving the request, an end device can use its own private key $D_{sk}$ to decrypt the payload and extract the request information. Then the device can independently determine whether the user should be authorized to have the corresponding permission. 


Now having approved access to device capabilities and attributes, $Alice$ develops a program using the interfaces provided by WASI-SN and sends it via the topic $/resources/device01/BME280$ with a signature signed by her private key $U_{sk}$ issued by PKG center. The signed message is composed of the Alice's UII and the payload decryption key $D'_{p}$ encrypted by the device's public key $D_{pk}$ defined above. 
The corresponding decryption key of the payload will be derived as $D'_{p}$ which is encrypted using device's public key $D_{pk}$ and be put inside the signature. When the Wasm runtime running on the specified device receives the payload, it will check the signature to ensure a valid requester, and decrypts the payload using its private key $D_{sk}$ to obtain the Wasm module. Before and during execution, the runtime checks all interface function calls used in the module to ensure that accesses to capabilities and attributes are valid and approved, thus enabling \textit{interactive} and \textit{secure} sensor network deployments.



\subsection{Key Revocation}
\subsubsection{Complete Subtree Broadcast Encryption}
For the key revocation scheme, we directly use the JEDI scheme that supports both \textit{Key Expiry} and \textit{Immediate Key Revocation}~\cite{kumar2019jedi}. The basic idea is to manage a Complete Subtree (CS) for different entities. The objective of the requester hierarchy structure previously introduced is to aid in the construction of the Complete Subtree. Each user corresponds to a leaf node, where all parents represent different domains inside the uniform identity identifier (UII), and each node contains separate WKD-IBE keypairs for its domain. Each user thus has the all secret keys for all the nodes belonging to the root-leaf path. In order to encrypt a message that is only decryptable for a subset of the users, the encryption key should be chosen from the collection of subtrees which include all the necessary leaves. Then the message is encrypted multiple times by using different public key of the root of each subtree.

\subsubsection{Immediate Revocation}
When a key is revoked, the issuer (i.e. end device) searches the complete subtree to find the range of the leaves to be revoked. The end device new encryption is performed by selecting the subtree of the remaining valid leaves and using the public key of each of their ancestor nodes. After a key is revoked, all the eliminated leaf node users are no longer able to decrypt the published messages. The same method applies to \textit{Key Expiry}, but instead each device monitors for expired leaf nodes before perfoming key revocation.


\section{Evaluation}
We evaluate the performance of the Wasm extended interface and access control overhead of our scheme by applying it to a real embedded device testbed, thereby comparing our WebAssembly implementation with native code. All the experiments are run on NRF52840-DK boards with Cortex-M4F@64 MHz MCU under the Armv7E-M architecture and 256KB SRAM. The board use IEEE 802.15.4 as the link and physical layer specification to communicate. In the results reported below, all the data labelled as native indicates the execution of the program run in the Zephyr RTOS; those not labelled as native are the results of running Wasm module with our customized runtime based on Wasm-micro-runtime(WAMR) from Intel inside the Zephyr RTOS. To setup the experiments, we choose Rpi 3B+ with Cortex-A53@1.4GHz CPU as the gateway which is connected to a broker running on a PC with Intel i7-6700@4 GHz CPU. To setup the experiments, the connection between the end device and gateway is in the hybrid mode to avoid a single point of failure. In order to using IP as the network protocol between the embedded device and gateway, we choose OpenThread to form up a Thread network to manage the connection with NRF52840-dongle as the Network Co-processor(NCP). The sensor used in the following experiments targets at BME280 sensor. The experiment setup is shown in Figure \ref{setup} below. 

The code size and lines of code for the evaluation programs developed in (1) a C program compiled to WebAssembly, and (2) the native C application are presented in Table \ref{tab:codesize}.
\begin{table}
\begin{tabular}{|c|c|c|}
\hline  
Applications&Wasm module&Native\\
\hline  
Access Sensors&10 LOC (357B)&23 LOC (893B)\\
\hline  
MQTT-SN client&16 LOC (523B)&97 LOC (3,003B)\\
\hline  
\end{tabular}
\caption{Code size for evaluated Wasm modules vs native}
\label{tab:codesize}
\vspace{-3mm}
\end{table}
Unlike native applications, most of the WebAssembly functionality is handled by the Wasm runtime, thereby easing the coding and maintenance burden on the application developer. Furthermore, as Wasm modules are portable across devices, they do not require a custom implementation, thereby enabling code reuse. 
\begin{figure}[htbp] 
\centering 
\includegraphics[width=\linewidth]{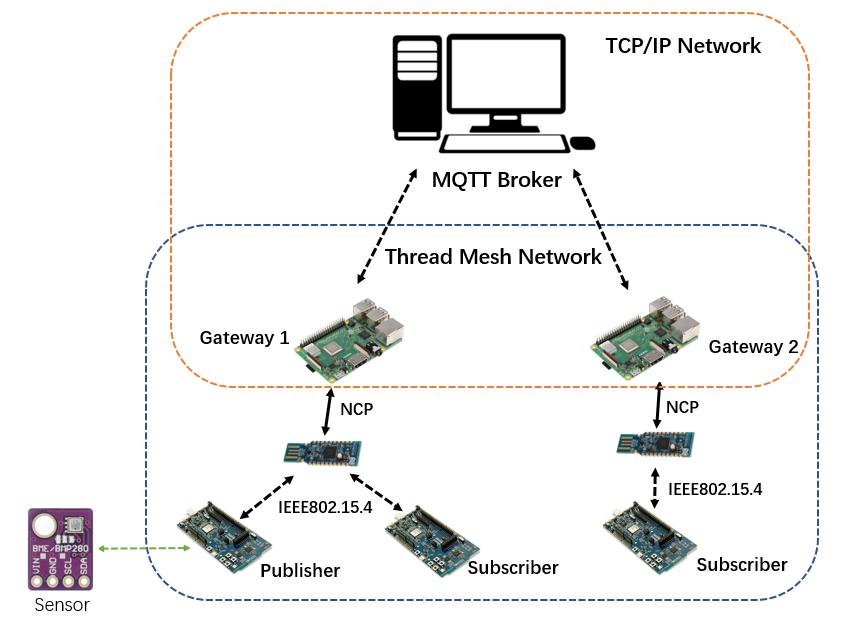}
\caption{Experiment Setup} 
\label{setup}
\end{figure}
\subsection{WASI Sensor Interface Overhead}
\textbf{Memory Footprint}: We first measure the additional memory footprint overhead introduced by using Webassembly as the target binary format. To emphasize the memory consumption overhead of the sensor interfaces, the main function of the application only contains the sensor interfaces defined in Section 3.3. In this first metric, the application initiate the sensor and attempt to retrieve the temperature value only once. The results are shown in Table \ref{tab:wamrmemory}.
\begin{table}
\begin{tabular}{|c|c|c|}
\hline  
Environment&Compilation&Runtime\\
\hline  
Native&125,162B&\textit{N/A}\\
\hline  
WASI-SN&131,500B&10,189B\\
\hline
WASI-SN + MQTT-SN&212,176B&10,189B\\
\hline  
\end{tabular}
\caption{Memory consumption of the WASI-SN WAMR implementation (also with MQTT-SN) compared to native. Applications can be optionally precompiled to Wasm to eliminate compilation overhead.}
\label{tab:wamrmemory}
\vspace{-4mm}
\end{table}

WASI-SN introduces a 5\% memory overhead during compilation when compared to native compilation. Given the 256KB SRAM size of the board onto which they are deployed, this is likely to be a generally acceptable overhead. In addition, given that Wasm is intentionally designed as a portable bytecode, applications can be precompiled on other host machines to eliminate compilation memory requirements entirely.
These results use a 100kB memory pool allocated to the runtime for compilation. For a fair comparison, we also allocated 100KB statically to the native application.

During module execution, WASI-SN only consumed ~10kB, a relatively small overhead compared to the static pre-allocated overhead during compile time. The dynamic memory footprint measured at runtime remained at around 10437B. We are unable to report the usage of the native program due to the nature of executing the native program on the Zephyr RTOS, so the native results are not reported.
The MQTT-SN library introduces a notable memory footprint during compilation, mostly due to the code size of the library itself; however, the addition of this library poses no effect during runtime for modules that do not leverage MQTT-SN in their program. 

\begin{figure}[ht] 
\centering 
\includegraphics[width=0.95\linewidth]{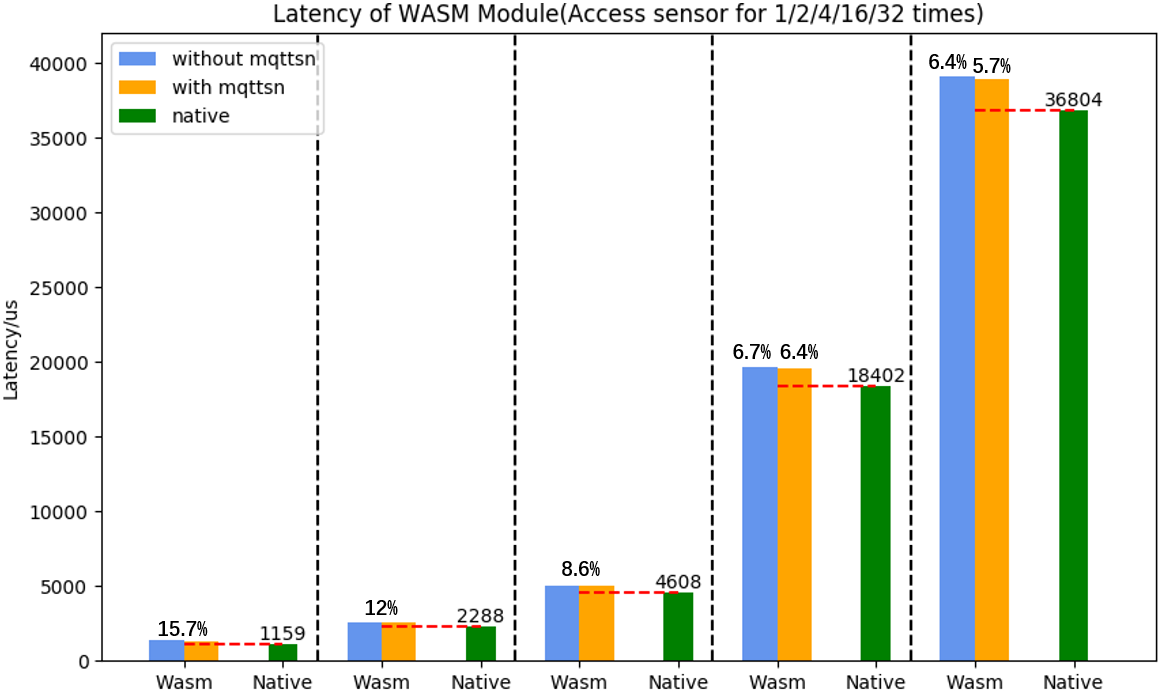}
\caption{Latency of WASI-SN compared with native when accessing sensors a varying number of times} 
\label{Latency_AS}
\end{figure}

\textbf{App Latency}: Next, we measured the latency when accessing sensors through the Wasm interface with respect to native application. We profile the latency by retrieving the temperature value from BME280 sensor a variable number of times to indicate the fixed startup costs in loading the WAMR runtime. Results are shown in Figure \ref{Latency_AS}.
The latency overhead when invoking sensor interfaces diminishes as the sensor is accessed repeatedly; after 32 subsequent accesses, latency differences stabilize to around a ~6\% overhead. In addition, the overhead latency between the runtime with and without the MQTT-SN library is nearly identical. This is expected, as the Wasm program does not use the MQTT-SN interface in this evaluation. 

\begin{figure}[ht] 
\centering 
\includegraphics[width=0.95\linewidth]{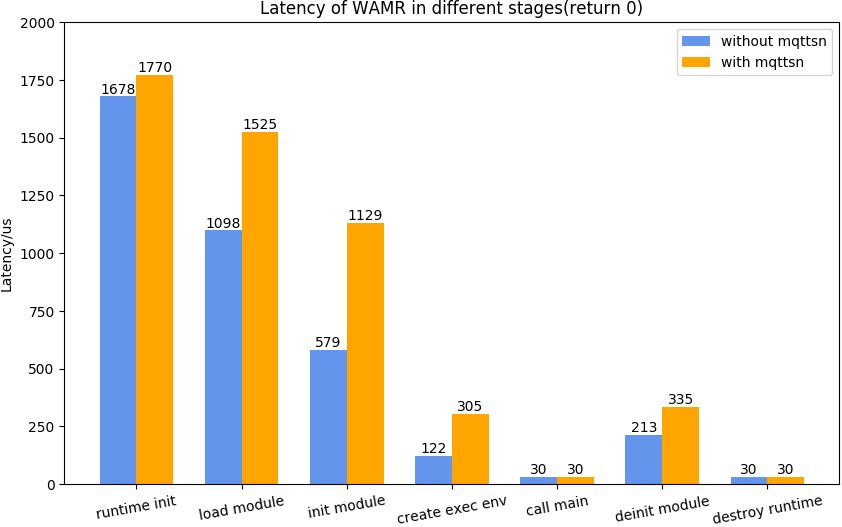}
\caption{Latency analysis of the WASI-SN runtime lifecycle. All stages except for the \textit{main} can be reused by the runtime for subsequent modules. Stage descriptions are provided in Table \ref{tab:wasmruntimelifecycle}.} 
\label{Latency_ret0}
\end{figure}

\textbf{Wasm Runtime Latency}: To further isolate the startup overhead of the WASI-SN runtime, we performed a lifecycle analysis of the various stages in runtime initialization, execution, and destruction. Each stage described in Table \ref{tab:wasmruntimelifecycle} was measured to provide a compherensive review of the independent elements. It is important to note that most of these stages are only incurred when the runtime is first initialized; afterwards, multiple modules can be executed subsequently. The main function used was a trivial main function that simply returned. Results are shown in Figure \ref{Latency_ret0}.

\begin{table}
\caption{The Wasm runtime lifecycle}
\label{tab:wasmruntimelifecycle}
\begin{tabular}{|c|c|}
\hline  
Stage&Functionality\\
\hline  
{\textbf{runtime init}}&{Initialize runtime and allocate memory}\\
\hline  
{\textbf{load module}}&{load Wasm module}\\
\hline  
{\textbf{init module}}&{initialize functions for modules}\\
\hline  
{\textbf{create exec env}}&{create sandbox for modules}\\
\hline  
{\textbf{call main}}&Execute a module\\
\hline  
{\textbf{deinit module}}&destruct variables sequentially\\
\hline  
{\textbf{destroy runtime}}&terminate runtime, clean up memory\\
\hline 
\end{tabular}
\end{table}


\begin{figure}[t] 
\centering 
\includegraphics[width=\linewidth]{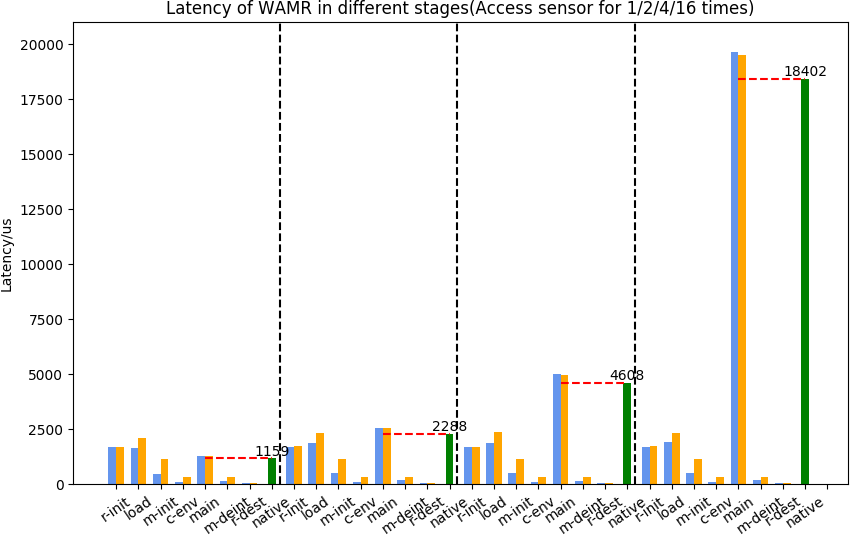}
\caption{Latency analysis of the WASI-SN runtime lifecycle for a variable number of sensor accesses. As the Wasm module increases in execution runtime, the fixed costs of runtime and module initialization are diminished.} 
\label{Latency_AS_stages}
\end{figure}

The majority of startup costs are in the stages before loading the Wasm module with an associated sandbox environment (i.e. ``booting'' the runtime). The addition of MQTT-SN libraries added a significant overhead in these upfront costs, likely due to the need for processing additional symbols and loading the relevant libraries (e.g. network interface). Nevertheless, the MQTT-SN extensions incur not additional latency during module execution, so long as the module does not access these particular extensions. Figure \ref{Latency_AS_stages} presents the complete breakdown of the various runtime stages as the number of sensor accesses increase. As we can see, the startup costs do not change as the module increases in complexity or execution runtime, and the WASI-SN extensions provide a similar 6\% overhead to native execution. After potentially multipl execution of a module, the runtime only needs to deinitalize the module without destroying itself to begin running another module, similar to the common main loop of embedded applications.


\subsection{WASI Network Interface Overhead}
\begin{figure*}[htbp]
\centering
\begin{minipage}[t]{0.33\linewidth}
\centering
\includegraphics[width=\linewidth]{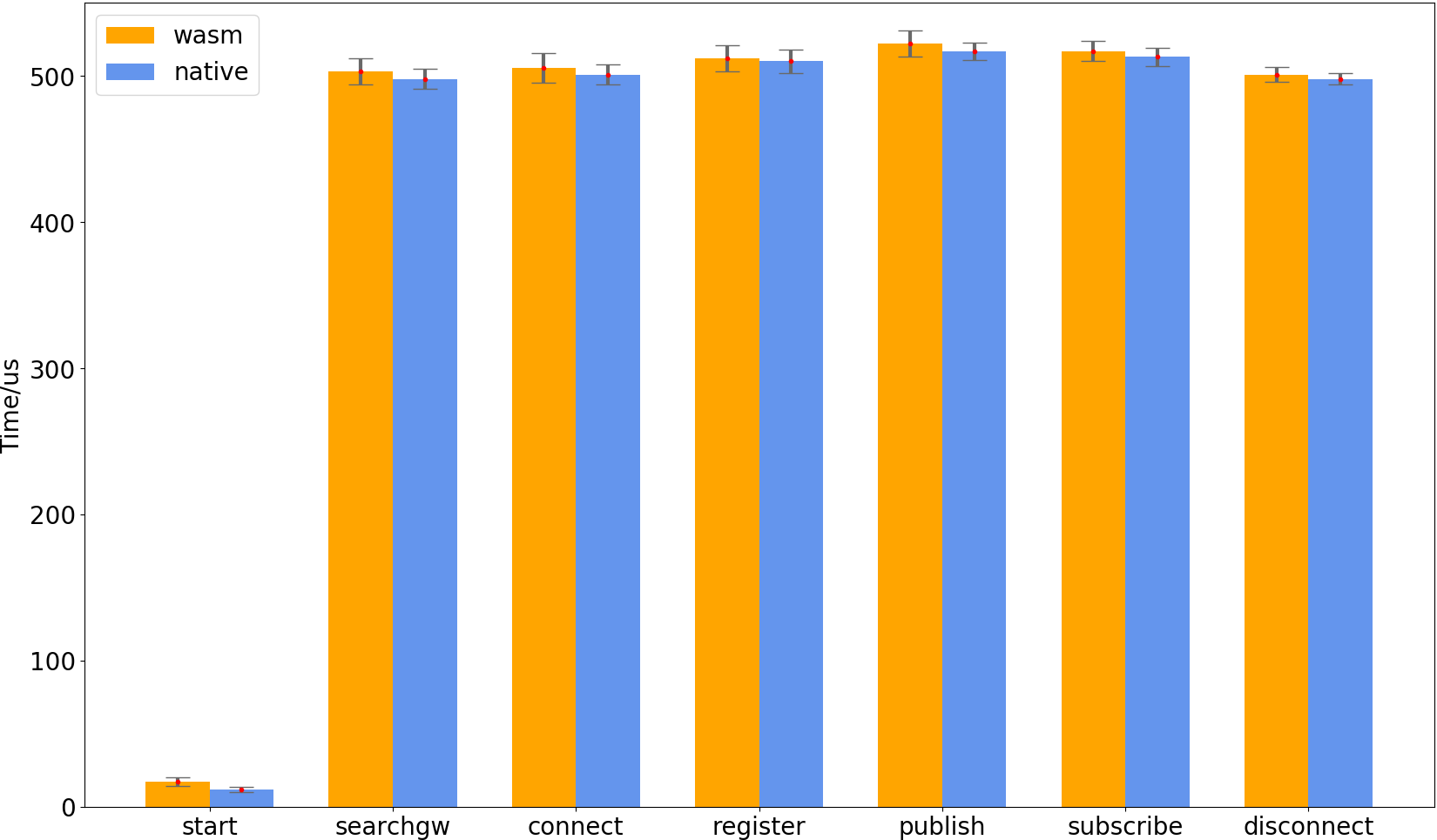}
{(a) Latency of executing different MQTT-SN network interface requests. }
\end{minipage}
\begin{minipage}[t]{0.33\linewidth}
\centering
\includegraphics[width=\linewidth]{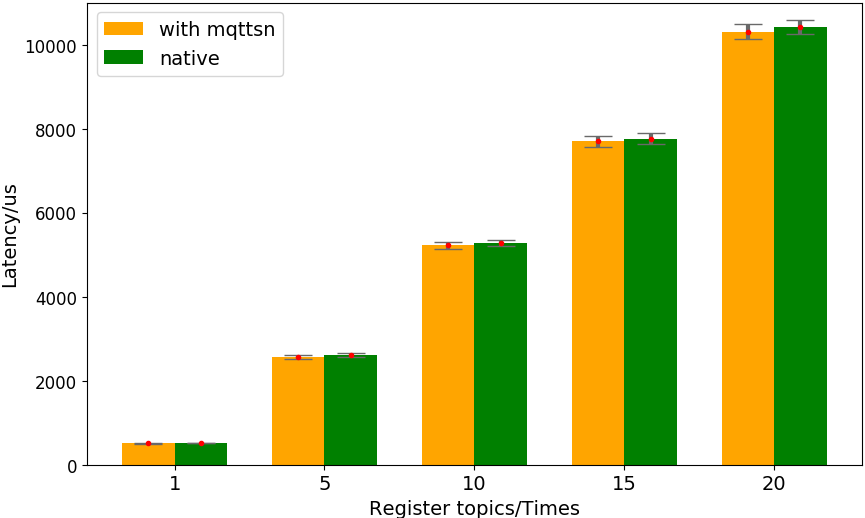}
{(b) Latency of registering an increasing number of topics. }
\end{minipage}
\begin{minipage}[t]{0.33\linewidth}
\centering
\includegraphics[width=\linewidth]{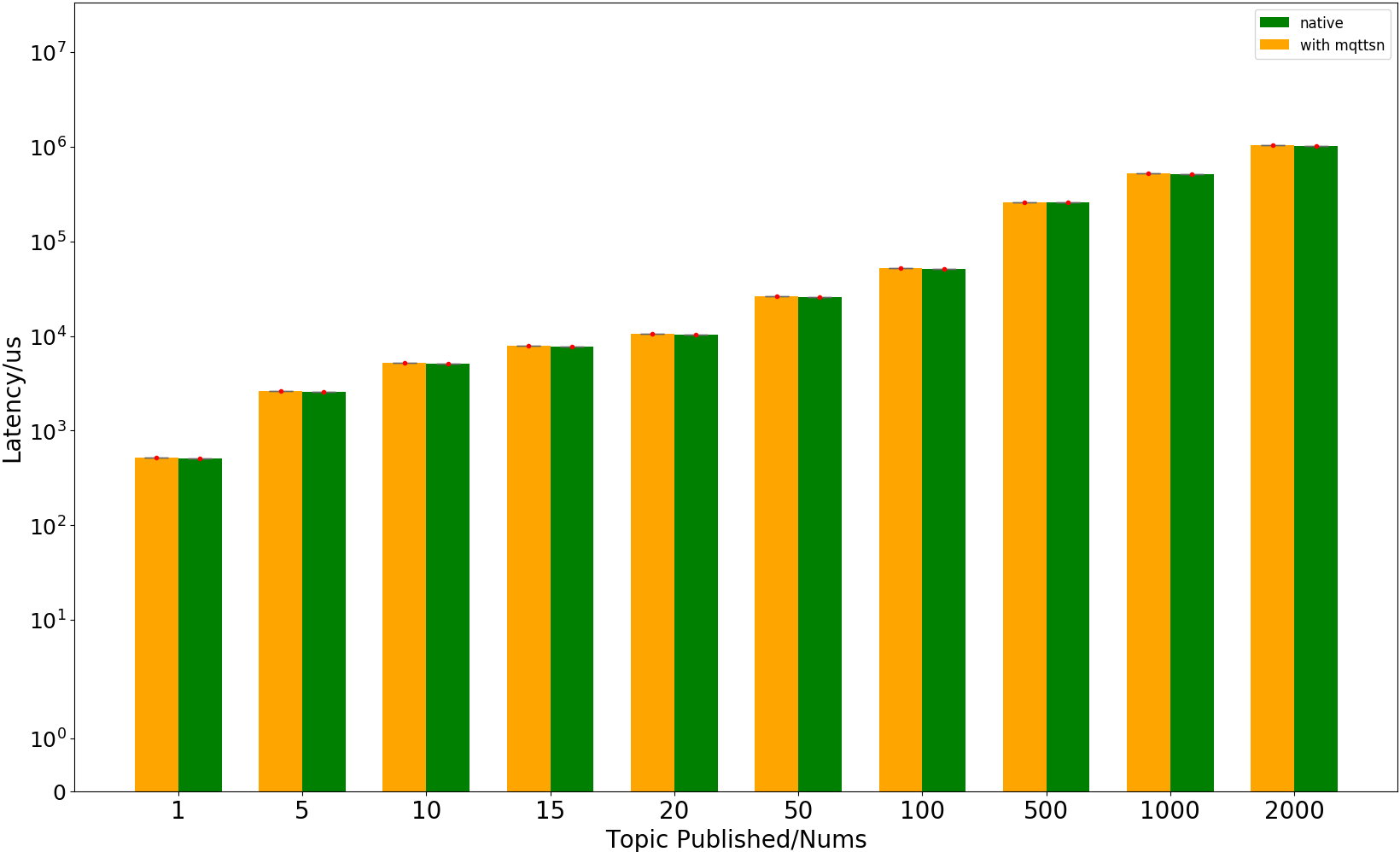}
{(c) Latency of publishing temperature values an increasing number of times.}
\end{minipage}
\caption{Latency measurement of various MQTT-SN network interface operations in WASI-SN}
\label{fig:mqttsneval}
\end{figure*}

Next, we performed an end-to-end latency evaluation of the MQTT-SN interface compared with native interface. To determine end-to-end latency, we measured the time between invoking the interface and receiving an acknowledgement from the gateway. We selected end-to-end latency as it provides practical results on the impact on WASI-SN from a developer's perspective. We first measure the latency when using the different network interfaces defined in Section \$4. The results are presented in Figure \ref{fig:mqttsneval}.

%
The latency involved in running the \textbf{start} interface is trivial as it does not involve any network transmission. Instead, it simply starts a MQTT-SN client with in the WebAssembly sandbox at a specific port. When executing the other network interfaces which require sending packets to the gateway, the latency of the WASI-SN implementation is nearly the exact same as native. This is primarily due to the fact that network delays dominate the latency results. 
Regardless of the network interface request, the overhead added by WebAssembly is negligible. 

Figure \ref{fig:mqttsneval}b/c present more common scenarios where applications may register across multiple topics and publish multiple messages. The latency involved in registering multiple topics is linear with the number of topics, as expected, and is less than 1\% different from native. Even when an application registers for 20 topics, the resulting latency is still around 1s, which is a generally acceptable latency; once registered, multiple messages can be pushlished without requiring additional registration. As the number of messages published increases, we see similar linear scaling with negligible latency overhead.

In summary, there is a minimal difference in performance when leveraging our introduced implementation of WASI-SN with the associated MQTT-SN extensions in terms of both the sensor access and network interface overhead. As such, using Webassembly as a SFI solution for networked and embedded systems provides a minimal impact on application performance while introducing a number of added security and portability benefits.

\parskip=0pt

\section{Related Work}
\textbf{Webassembly for IoT device}: Webassembly has drawn significant attention from both industry and academia. The trend to use Wasm as portable binary target for embedded device has become a recent topic of focus~\cite{napieralla2020considering,ewasm}. Elliott Wen et al.,~\cite{wen2020wasmachine} designed a lightweight OS for running Webassembly on the IoT device with a Rust based kernel to secure the Wasm application. Jacobsson et al.,~\cite{jacobsson2018virtual} exploits the possibility to run Webassembly across multiple embedded device with different architecture. Company like Intel~\cite{Intel:2019} also extends the security feature by creating Wasm runtime to extend the security feature of Wasm binary format and enforce memory check in the runtime. Alexandru~\cite{radovicisecure} then proposed a Webassembly framework to not only run Webassembly on IoT device but also enable the dynamic updates to correct the flaws in Wasm binary.  

\textbf{Security over MQTT/MQTT-SN}: Many existing researches have introduced various security schemes over MQTT-SN to mitigate the problem of lacking end-to-end encryption and access control. Ousmane et al.,~\cite{sadio2019lightweight} proposed ChaCha20-Poly1305 Authenticated Encryption with Associated Data(AEAD) as the lightweight encryption scheme over the data transmitted through MQTT/MQTT-SN. Amaran, M., et al.,~\cite{amaran2018lightweight} measures the performance of different encryption scheme over MQTT-SN such as LBLOCK, AES and KLEIN and drew a conclusion that LBLOCK performs the best overall. Muhammad, Ahsan, et al.,~\cite{muhammad2019onem2m} implements a secure MQTT binding following the oneM2M standard architecture to provide interoperability and secure communication channel among IoT devices. In other works, Sochor et al.,~\cite{sochor2020exploiting} exploit the attack surface of MQTT-SN protocol. They successfully perform distributed reflection denial-of-service(DRDoS) over MQTT-SN for embedded device and proposed the corresponding countermeasures. Singh et al.,~\cite{singh2015secure} proposes a secure version of MQTT and MQTT-SN(SMQTT/SMQTT-SN) by leveraging Key/Ciphertext Policy-Attribute Based Encryption (KP/CP-ABE) to secure data during transmission.

\textbf{Identity Based Encryption over IoT}: With the rapid developement of IoT devices, the community has explored opportunities in IBE as an encryption scheme for IoT networks for a lightweight solution to maintain security guarantees. Yang Liu et al.,~\cite{yang2013iot} build a network security architecture for IoT by combining IBE and PKI/CA to enable secure authentication and data encryption through the network. Jebri et al.,~\cite{jebri2015efficient} proposed a secure generic model for IoT and wireless sensor network to guarantee anonymity, secrecy between devices with the help of IBE and Pseudonym Based Encryption (PBC). Kumar et al.,~\cite{kumar2019jedi} recently proposed a many-to-many end-to-end encryption protocol based on wildcarded IBE(WKD-IBE)~\cite{abdalla2011wildcarded} and improve the design by allowing fast key delegation and expiry/immediate revocation scheme, which is leveraged in our MQTT-SN access control approach.
\section{Conclusion}
In this paper, we proposed WASI-SN, the first implementation of an extension to the WebAssembly System Interface enabling Wasm applications running on embedded device to have a unified sensor interface definition with networked access control. The proposed interfaces follows previously defined capability-based schemes used to categorize, configure, and access sensors. In addition, we implement the first MQTT-SN library to work with Webassembly on top of Zephyr and design an end-to-end wildcard-based identity-based encryption(WKD-IBE) scheme to protect the privacy of data without involving broker and gateway as part of the trust chain. Overall, our framework provides a chance to improve the portability and security of the embedded application, particularly those that require sensor and network access, with close to native overhead. This design of combining Webassembly with embedded device using well defined interfaces can provide a next-step in the progress toward running secure and portable applications across heterogeneous platforms, thereby hoping to advance the next generation of IoT and edge computing. 
We firmly believe that Webassembly will soon become the new dominant binary format in the embedded world of the future.
\bibliographystyle{ACM-Reference-Format}
\bibliography{references.bib}
\end{document}